\documentclass[a4paper,onecolumn,11pt,unpublished]{quantumarticle}
\pdfoutput=1
\usepackage[utf8]{inputenc}
\usepackage[english]{babel}
\usepackage[T1]{fontenc}

\usepackage[usenames,dvipsnames]{color}
\usepackage[hyperfootnotes=false]{hyperref}
\hypersetup{
	colorlinks,
	citecolor=quantumviolet,
	linkcolor=quantumviolet,
	urlcolor=quantumviolet}

\usepackage{tikz}
\usepackage{lipsum}
\usepackage[numbers,sort&compress]{natbib}

\usepackage{algorithmicx}
\usepackage[linesnumbered,ruled,vlined]{algorithm2e}
\usepackage[noend]{algpseudocode}
\usepackage{mathtools}
\usepackage{graphicx}
\usepackage{amssymb}
\usepackage{amsmath}
\usepackage{amsthm}
\usepackage{amsfonts}
\usepackage{color}
\usepackage{xcolor}
\usepackage{braket}
\usepackage{soul}

\usepackage{subfigure}
\usepackage{booktabs}
\usepackage{hyperref}

\SetKwInput{KwInput}{Input}                
\SetKwInput{KwOutput}{Output}              

\DeclareMathOperator{\tr}{\mathrm{tr}} 
\newcommand{\Qset}[1]{\mathbb{Q}_{#1}}

\definecolor{CCTLABgreen}{RGB}{0,128,0}

\begin{document}

\title{Matrix-Completion Quantum State Tomography}

\author{Ahmad Farooq}
\orcid{0000-0002-6090-9553}
\author{Junaid ur Rehman}
\orcid{0000-0002-2933-8609}
\author{Hyundong Shin}
\orcid{0000-0003-3364-8084}
\affiliation{Department of Electronics and Information Convergence Engineering, Kyung Hee University,\\ 1732 Deogyeong-daero, Yongin-si, Gyeonggi-do 17104, Korea}
\email{hshin@khu.ac.kr}
\homepage{http://cctlab.khu.ac.kr}
\thanks{(corresponding author)}
\maketitle

\begin{abstract}
	The deployment of intermediate- and large-scale quantum devices necessitates the development of efficient full state tomographical techniques for quantum benchmarks. Here, we introduce a matrix filling-based method for tomography of pure quantum states, called the matrix-completion quantum state tomography.  This method requires only $2n + 1$ local Pauli operators and minimal post-processing  for  $n$-qubit states. Numerical results show that our method is highly efficient on superconducting real quantum devices and achieves better fidelity estimates of multiqubit quantum states as compared to contemporary pure state tomography methods. These desirable features of the matrix-completion quantum state tomography protocol make it suitable for the benchmarking of intermediate- and large-scale quantum devices dealing mainly with pure quantum states.
\end{abstract}

\section{Introduction}
Noisy intermediate-scale quantum (NISQ) devices consisting of hundreds of qubits will be available soon \cite{Pre:18:Quantum}. Several technologies for these NISQ devices are being pursued such as superconducting quantum circuits \cite{KMTTBCG:17:Nat}, trapped ions \cite{ZPHKBKGGM:17:Nat}, quantum dot \cite{LD:98:PRA}, cold atoms \cite{BSKLOPCZEG:17:Nat}, and photonic platforms \cite{RZBB:94:PRL}. These NISQ computers are expected to perform tasks that surpass the capability of the most powerful classical computers available today \cite{Ter:18:NPh}. However, noisy quantum gates and decoherence limits prevent exhibiting a clean enough improvement over the existing current classical computing devices for complex algorithms. With the advent of the NISQ era, we come across a paradox: How will we validate a quantum device confirming that it is producing the desired result that it is designed for? 
For characterization, certification, and benchmarking \cite{JHWRMPCK:20:NRP} of these noisy devices, quantum state tomography comes into play, which is the gold standard for the reconstruction of a quantum state \cite{TM:20:ARCMP}. 
Reconstruction of the quantum state from measurements on the replica of an unknown given quantum state is termed as the quantum state tomography. Quantum state tomography generally consists of a two-step process: the first is collecting the data from optimized measurement design on quantum systems; the second is the classical post-processing of the gathered data. Quantum state tomography problem becomes intractable due to an exponential growth of the system size with increase in the number of qubits under consideration \cite{SKMRBAFW:11:PRL,LMR:14:NPh}.

A conventional state standard tomography employs $d^{2}$ or more measurement settings, where $d=2^{n}$ for $n$ qubits \cite{BDPS:99:PRA,OWV:97:PRA,QQHLDXG:13:SR,Rob:10:NJP}. 
A simple counting argument suggests that only $O\left(d\right)$ measurement settings are sufficient to perform the tomography of pure quantum states \cite{DM:97:PLA,AW:99:JPAMG}. Compressed sensing and low-rank matrix recovery from sparse matrix algorithms have been developed to achieve this bound \cite{SRMRBMTRE:17:QST,KRT:17:ACHA,BKGL:17:njpQI}. In \cite{KF:89:AP,AS:10:PRL}, the authors proposed $d+1$ mutually unbiased bases measurements for \emph{full} quantum state reconstruction. The states with special properties, for instance, matrix product states \cite{CPFSGBCPL:10:NC} or permutationally invariant quantum states \cite{TWGKSW:10:PRL} also lead to a significant reduction in the number of quantum measurement settings. The pure quantum state can also be retrieved with only five and three measurement settings given in \cite{GCEGXLD:15:PRL,CHKST:16:EPL,ZPMCLD:20:PRAp}. Recently noise resilient and robust self-guided quantum state tomography is proposed \cite{RQKFWR:21:PRL,Fer:14:PRL,CFP:16:PRL}.  Five basis measurement has the highest performance among all the non-adaptive quantum state tomography algorithm.

All the aforementioned tomography algorithms employ entangled bases in the measurement setting. For the implementation of entangled bases on NISQ devices, we apply sets of unitary transformations followed by the computational basis measurement. The controlled-NOT (CNOT) gate is an essential transformation requierd for implementing these sets of entangled measurements. The implementation of CNOT gate on quantum devices introduces a larger error due to experimental difficulties. 
Moreover, the circuit depth is quite high for the large number of qubit systems, which results in the decoherence of qubits.  
Hence, these entangled bases significantly reduce the performance of state estimation task. The introduction of local basis measurement for state reconstruction can be a massive catalyst for the benchmarking of quantum devices.  For this purpose, local basis measurements are introduced in \cite{PZD:21:arXiv}. They have demonstrated the task of tomography of pure quantum state by using $kn+1$ measurement bases, where $k \geq 2$. Their algorithm is based on solving several systems of linear equations. The required number of bases increases ($k > 2$) if the determinant of the linear system of equations vanishes. The performance is quite low as compared to the numerical results of entangled measurement algorithms. 

In this paper, we present a matrix filling-based quantum state estimation algorithm for pure states, called the matrix-completion quantum state tomography. Our algorithm utilizes only $2n+1$ local Pauli measurement bases for an $n$-qubit state. We first experimentally obtain a sufficient number of elements of the density matrix such that a matrix-completion algorithm for rank-1 matrices can be employed. We then fill the missing entries by taking any $2 \times 2$ submatrix in the density matrix using the rank one constraint. Simulation and experimental numerical examples demonstrate that the performance of our algorithm is notably high as compared to the existing algorithms. The matrix-completion quantum state tomography also has both computational complexity and memory storage low which is the order of  $O\left(d^{2}\right)$. 
\section{Method}
In this section, we outline our protocol for tomography of pure quantum states by matrix filling. Our technique requires $2n +1$ local measurement settings for an $n$-qubit system. The main ingredient of this technique is the algorithm for completing rank-1 matrices \cite[Section~IV.8]{Str:19:well}. This algorithm makes use of the fact that all columns/rows of a rank-1 matrix are multiples of each other. Alternatively, determinants of all $2\times 2$ submatrices must be zero. We first experimentally obtain a sufficient number of entries of the density matrix of a pure state. Then, we employ the matrix completion to obtain the complete density matrix of our state of interest.

\begin{figure}[t!] 
	\centering     
	\subfigure[Success]{\label{fig:a}\includegraphics[width=35mm]{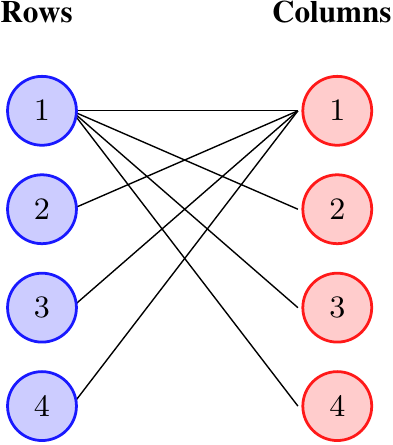}}
	\hspace{1.5cm}
	\subfigure[Success]{\label{fig:b}\includegraphics[width=35mm]{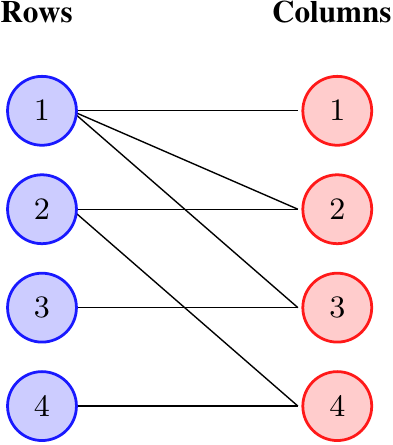}}
	\hspace{1.5cm}
	\subfigure[Failure]{\label{fig:c}\includegraphics[width=35mm]{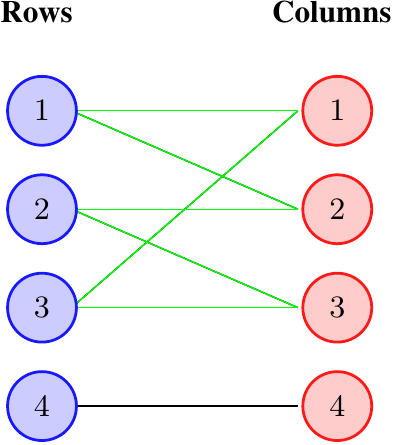}}
	\caption{ Criteria to successfully construct the rank-1 $d \times d$ matrix from the $2d-1$ entries. Edge connection between $i$th row node and $j$th column node represent the matrix element entry $a_{ij}$. By knowing only $2d-1$ entries of the matrix, we can construct the whole rank one matrix of dimension $d \times d$ when the edges connection between row-column tree do not make any cycle. The tree diagram is shown for $4 \times 4$. In $\bold{(a)}$ and $\bold{(b)}$, none of the edges make the complete cycle while $\bold{(c)}$  shows that the edges connection form a complete loop.    }\label{fig1}
\end{figure}

Filling the missing entries of a rank-1 matrix is directly connected to the problem of finding cycles in a graph \cite{Str:19:well}. The rank-1 matrix of dimension $d\times d$ can be fully reconstructed by knowing only $2d-1$ entries of the given matrix. These $2d-1$ elements form a spanning tree in the row-column graph.  The successful reconstruction of missing entries of the given matrix depends on the tree reaching all nodes with no loop. The tree with a loop in the row-column graph does not guarantee rank-1 matrix construction. If we know the $2d-1$ entries of the matrix which do not form the cycle in the row-column tree graph, then we can construct successfully all the other entries of the matrix by utilizing the fact that every $2 \times 2$ determinant must be zero in the rank-1 matrix. The row-column graph is shown in Figure \ref{fig1}.

	\begin{figure}[t!]
	\centering
	\includegraphics[width = 1 \textwidth]{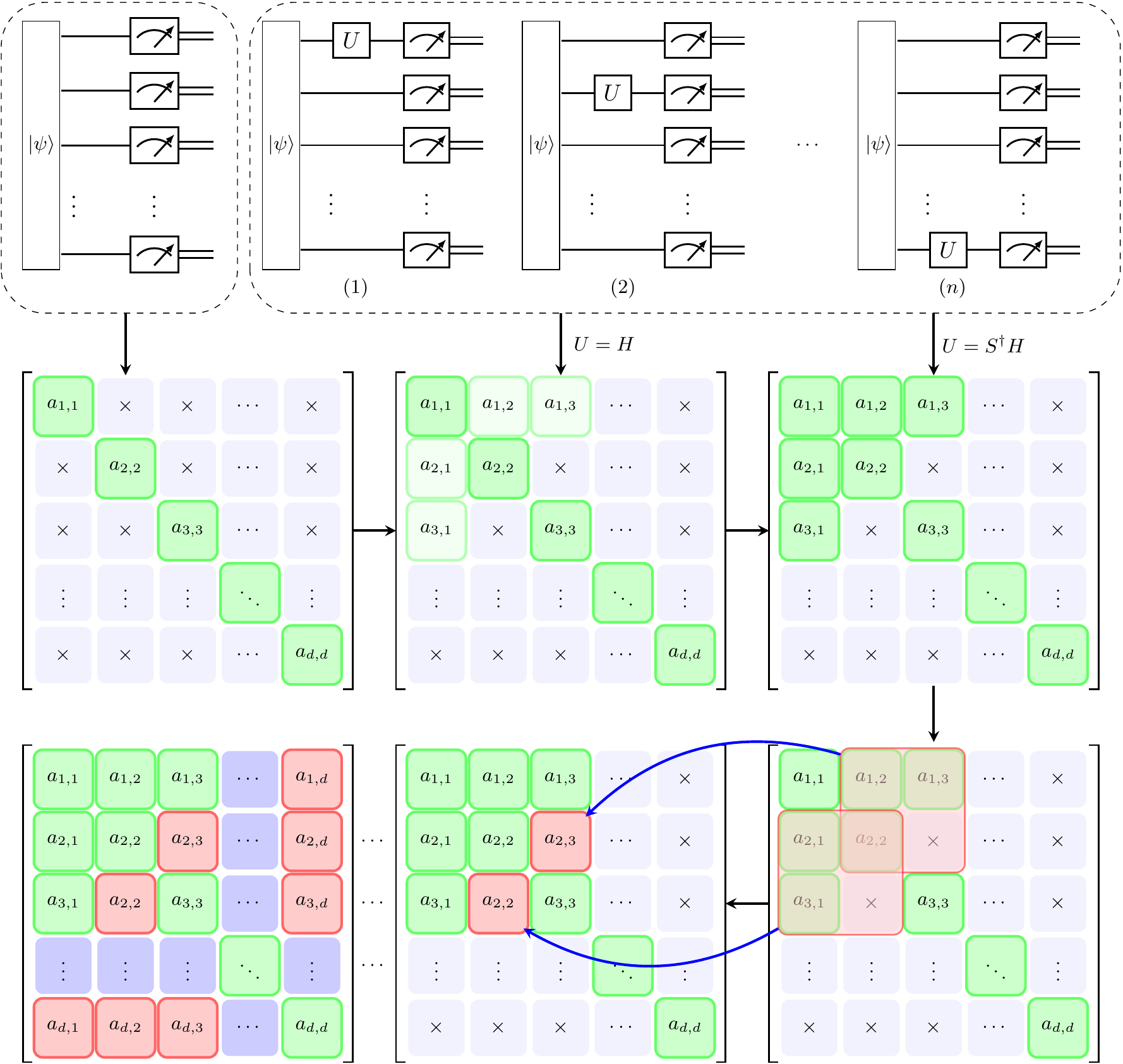}
	\caption{Overview of the matrix-completion quantum state tomography. The green color represents the entries that are obtained experimentally. The light green color shows only the real part of entries of the matrix. The red color represents the entries that are obtained by matrix completion. The dark blue color represents entries that can be obtained experimentally or from the algorithm.  We first obtain the diagonal entries of the rank-one matrix using computational-basis measurement. Using $U=H$ in the above circuit enables us to obtain the real component of the entries while using $U=S^{\dagger}H$ provides the Imaginary components of entries. Determinants of $2$ by $2$ submatrices of the light red color is used to find the missing entries in that submatrix.} \label{fig2}
\end{figure}

In our algorithm, we perform local Pauli measurements on each subsystem of multipartite pure quantum state.  
We first employ the computational basis measurement to obtain the diagonal elements of the unknown density matrix. 
Next, we measure $n$-qubit systems in Pauli measurement setting
\begin{align}
	\Lambda_{j}^{i}=I_{2}^{\otimes i-1}\otimes \sigma_{j}^{i}\otimes  I_{2}^{\otimes n-i}, \label{eq:1}
\end{align}
where $i=1,2,\cdots,n$, with $\sigma_{j}^{i} \in \left\{\sigma_1 ,\sigma_2 \right\}$ are Pauli $X$ ($\sigma_1$) and Pauli $Y$ ($\sigma_2$) matrices.

We begin by assuming that we have access to $N$ copies of a $n$-qubit multipartite unknown pure state $\rho = \ket{\psi}\bra{\psi}$.
In the first step of our algorithm we obtain the diagonal elements of $\rho$ by measuring $N/(2n + 1)$ of its copies in the computational basis of the $d$-dimensional Hilbert space. From the measurement results, diagonal elements of $\rho$ can be directly obtained as
\begin{align}
	\hat{\rho}_{i,i} = \frac{\left( 2n + 1\right)N_{i}}{N}, \text{ for }  i \in \Qset{d},
	\label{eq:diagonal}
\end{align}
where $\Qset{d} = \left\{0, 1, \cdots, d - 1\right\}$, $N_{i}$ is the number of times measurement result corresponding to $\ket{i}$ was obtained in the first step and $\hat{a}$ denotes an estimate of $a$. 

The off-diagonal element $\rho_{j, k}$ of $\rho$ can be obtained by measuring it in the eigenbasis of Pauli operators pair $\Lambda_{1}^{i}$ and $\Lambda_{2}^{i}$. This can be seen by first noticing that the eigenvectors of $\Lambda_{1}^{i}$ and $\Lambda_{2}^{i}$ are 
\begin{align}
	\ket{V_{1,\pm 1}^{2^{i}\nu+\zeta,2^i \nu+\zeta+2^{i-1}}}&=\frac{1}{\sqrt{2}}\left(\ket{2^{i} \nu+\zeta}\pm\ket{2^{i}\nu+\zeta+2^{i-1}}\right)\\
	\ket{V_{2,\pm 1}^{2^{i} \nu+\zeta,2^{i} \nu+\zeta+2^{i-1}}}&=\frac{1}{\sqrt{2}}\left(\ket{2^{i} \nu+\zeta}\pm \dot{\iota}\ket{2^{i}\nu+\zeta+2^{i-1}}\right),
\end{align}
where $\nu\in \Qset{\frac{d}{2}}$ and $\zeta\in \Qset{2^{i - 1}}$. 
\alglanguage{pseudocode}
\begin{algorithm}[t!]
	\small
	\caption{Pure State Quantum Tomography by Matrix Completion}
	\label{Algorithm:1}
	\KwInput{ $N$ copies of the $n$-qubit pure state $\ket{\psi}$.}
	\KwOutput{Estimate $\ket{\hat{\psi}}$ of $\ket{\psi}$}
	$\hat{ \rho}_{i,i}\leftarrow  \left|\braket{\psi|i}\right|^{2}$,\hspace{4.42cm} $\blacktriangleright$ Computational  basis measurement	\\
	\For{$i  \in 1 \rightarrow n $} 
	{
		\For{$x \in 1 \rightarrow 2$ }
		{
			$\Lambda_{x}^{i}\leftarrow I_{2}^{\otimes i-1}\otimes \sigma_{x}^{i}\otimes  I_{2}^{\otimes n-i}$ ,\hspace{1.3cm} $\blacktriangleright$ Pauli operators\\
			$\ket{V_{x,\pm 1}^{2^{i} \nu+\zeta,2^{i}\nu+\zeta+2^{i-1}}} \leftarrow \mathrm{eig}\left(\Lambda_{x}^{i}\right)$,\hspace{0.7cm} $\blacktriangleright$ where $\nu\in \Qset{\frac{d}{2}}$ and $\zeta\in \Qset{2^{i - 1}}$\\		
			$	\Pi_{x, \pm 1}^{j,k }\leftarrow \ket{V_{x,\pm 1}^{j,k}}\bra{V_{x,\pm 1}^{j,k}}$, \hspace{1.7cm} $\blacktriangleright$ where $j=2^{i} \nu+\zeta$ and $k=2^{i}\nu+\zeta+2^{i-1}$\\
			$\hat{p}_{x,\pm 1 }^{j,k}\leftarrow  \braket{\psi|\Pi_{x, \pm 1}^{j,k }|\psi}$,  \hspace{2.15cm} $\blacktriangleright$ Local Pauli measurements
		}
	}
	$	\hat{\rho}_{j,k}\leftarrow\frac{1}{2}\left(-\hat{p}_{1,-1}^{j,k}+\hat{p}_{1,1}^{j,k}-\iota \hat{p}_{2,-1}^{j,k}+\iota \hat{p}_{2,1}^{j,k}\right)$ ,  \hspace{0.1cm} $\blacktriangleright$ Density matrix elements entries\\
	$\hat{\rho}_{k, j} \leftarrow \hat{ \rho}_{j, k}^{\dagger}$, \hspace{4.85cm} $\blacktriangleright$ Hermitian symmetric \\
	$\hat{\rho}_{ij}\leftarrow\frac{\hat{\rho}_{ri}\hat{\rho}_{jr}}{\hat{\rho}_{rr}}$,\hspace{4.8cm} $\blacktriangleright$ Matrix completion of any $2 \times 2$ submatrix\\
	$\hat{\rho}\leftarrow \frac{\hat{\rho}}{\tr\left(\hat{\rho}\right)}$,\hspace{5.25cm} $\blacktriangleright$ Normalization\\
	$\ket{\hat{\psi}}\leftarrow \mathrm{eig}\left(\hat{ \rho}\right)$ ,\hspace{4.55cm} $\blacktriangleright$ Eigenvector with highest eigenvalue.\\
	return $\ket{\hat{ \psi}}$
\end{algorithm}
These orthonormal bases build a complete projector system with elements
\begin{align}
\Pi_{x, -1}^{j,k }& = \ket{V_{x,-1}^{j,k}}\bra{V_{x,-1}^{j,k}}\\
\Pi_{x, 1}^{j, k} &= \ket{V_{x,1}^{j,k}}\bra{V_{x,1}^{j,k}}.
\end{align}
 where $x\in \left\{1, 2\right\}$, $j=2^i \nu+\zeta$ and $k=2^i \nu+\zeta+2^{i-1}$. Let $p_{x, m}^{j, k}$ be the probability of obtaining the measurement outcome corresponding to the projector $\Pi_{x, m}^{j, k}$. Then, we have
\begin{align}
	p_{1, -1}^{j, k} &= \frac{1}{2}\left( \rho_{j, j} - \rho_{j, k} - \rho_{k, j} + \rho_{k, k}\right)\label{eq:p_1}\\
	p_{1, 1}^{j, k} &= \frac{1}{2}\left( \rho_{j, j} + \rho_{j, k} + \rho_{k, j} + \rho_{k, k}\right)\label{eq:p_2}\\
	p_{2, -1}^{j, k} &= \frac{1}{2}\left( \rho_{j, j} +\dot{\iota} \rho_{j, k} - \dot{\iota}\rho_{k, j} + \rho_{k, k}\right)\label{eq:p_3}\\
	p_{2, 1}^{j, k} &= \frac{1}{2}\left( \rho_{j, j} -\dot{\iota} \rho_{j, k} + \dot{\iota}\rho_{k, j} + \rho_{k, k}\right)\label{eq:p_4},
\end{align}
where $\rho_{j, k} = \braket{j|\rho|k}$ is the element of $\rho$ at $(j, k)$th index. From the hermiticity of $\rho$, we have $\rho_{k, j} = \rho_{j, k}^{\dagger}$. Using this fact, we get $\rho_{j, k} + \rho_{k, j} = 2 \mathrm{Re}\left\{ \rho_{j, k}\right\}$ and $\rho_{j, k} - \rho_{k, j} = 2 \mathrm{Im}\left\{ \rho_{j, k}\right\}$. Consequently, we can obtain $\rho_{j, k}$ as
\begin{align}
	\rho_{j,k}=\frac{1}{2}\left(-p_{1,-1}^{jk}+p_{1,1}^{jk}-\iota p_{2,-1}^{jk}+\iota p_{2,1}^{jk}\right).
	\label{eq:off_diagonal}
\end{align}
From $\rho_{jk}$, we can easily get $\rho_{k,j}=\rho_{j,k}^{\dagger}$.

The complete protocol for constructing the density matrix of a pure quantum state $\rho$ is as follows:
\begin{enumerate}
	\item  Measure $\rho$ in the computational basis to obtain the diagonal entries of the density matrix  through \eqref{eq:diagonal}.
	\item Obtain entries of the density matrix $\hat{\rho}_{j,k}$. These entries are obtained by measuring in all Pauli operators measurement setting of \eqref{eq:1} and using \eqref{eq:off_diagonal} as described above. 
		\begin{figure}[t!]
		\centering
		\includegraphics[width = 0.6 \textwidth]{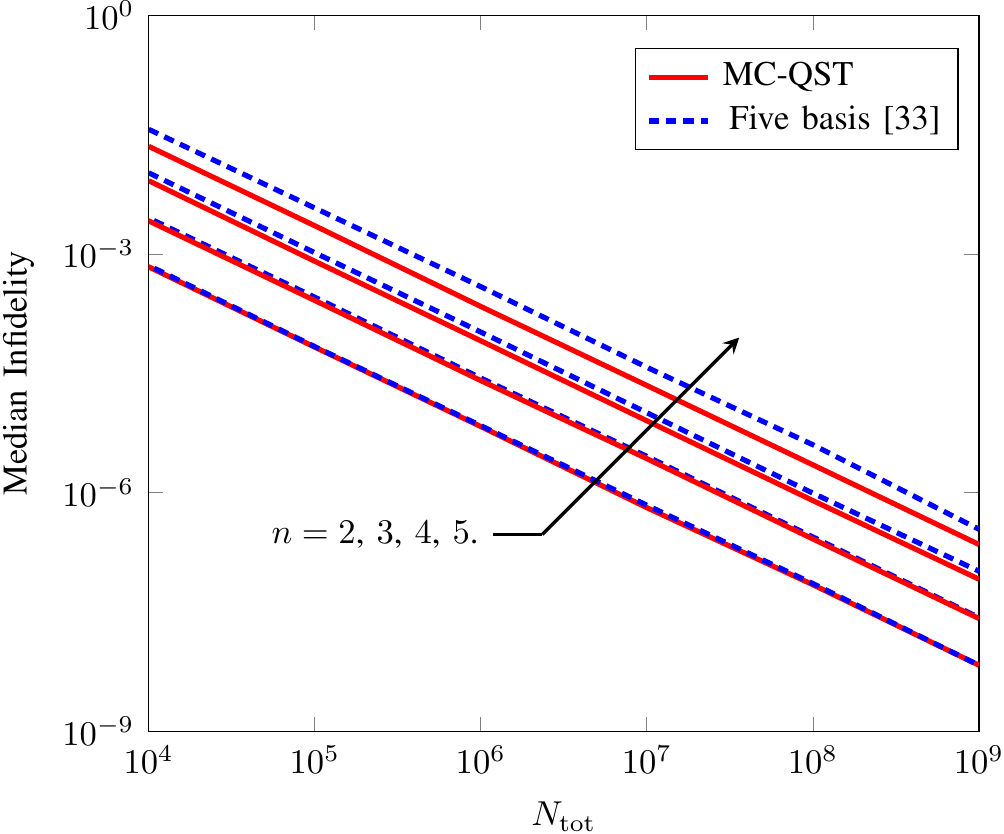}
		\caption{ Median Infidelities of $10^3$ randomly generated pure state from matrix-completion quantum state tomography (MC-QST) and modified five basis algorithm as a function of a number of total measurements. The performance of MC-QST improves as compared to the five basis algorithm as the number of qubits increases. } \label{fig10}
	\end{figure}

	\item  Since $\rho$ is pure, the corresponding density matrix has rank one. Thus, its remaining entries can be obtained by \cite[Section~IV.8]{Str:19:well}
	\begin{align}
		\hat{\rho}_{ij}=\frac{\hat{\rho}_{ri}\hat{\rho}_{jr}}{\hat{\rho}_{rr}}.
	\end{align}
	This is because, in any rank-one matrix, every $2 \times 2$ determinant must be zero. The computational complexity of this step is $O\left(d^{2}\right)$ which is the total computational complexity of our algorithm.
\end{enumerate}
A schematic diagram of matrix-completion quantum state tomography algorithm is shown in Figure~\ref{fig2} and complete procedure is outlined in Algorithm \ref{Algorithm:1}.

\section{Results}
We can certify the purity of our unknown quantum state using our algorithm without the need of any extra measurement. A density matrix $\rho$ is said to be pure if and only if the equation
\begin{align} \label{Purity}
	|\rho_{i,j}|^2=|\rho_{i,i}||\rho_{j, j}|,
\end{align}
holds for every $i,j=1,2,\cdots,d$. We can easily verify the purity of unknown quantum state from measured structure obtained in Figure~\ref{fig2} by employing \eqref{Purity} on any number of elements of the constructed density matrix.

To gauge the accuracy of state tomography, we use the common figure of merit, infidelity, which characterizes the distance between these states. Infidelity is defined as
\begin{align}
	1-F\left(\rho,\sigma\right)=1-\left(\tr\sqrt{\sqrt{\sigma}\rho\sqrt{\sigma}}\right)^{2} 
\end{align}
where $F\left(\rho,\sigma\right)$ is the fidelity between $\rho$ and $\sigma$.
\begin{figure}[t!]
	\centering
	\includegraphics[width = 0.6 \textwidth]{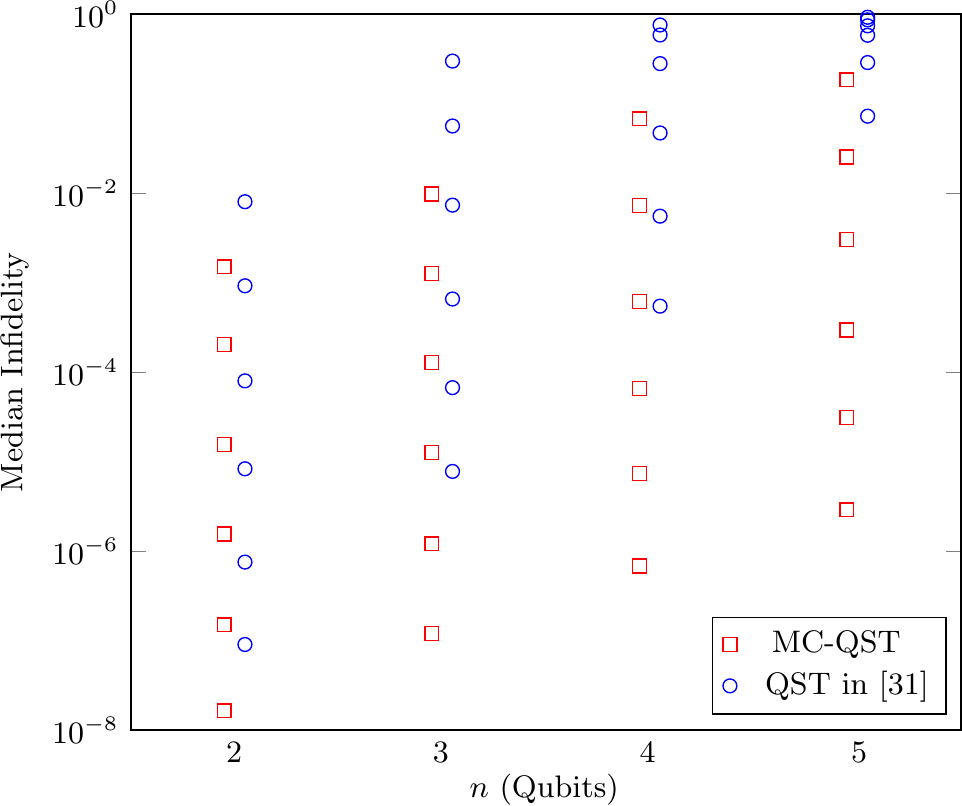}
	\caption{Median infidelity as a function of number of qubits by varying the measurement copies in each measurement setting. The median infidelities of $10^{3}$ Haar-random pure quantum states obtained from matrix-completion quantum state tomography (MC-QST) and algorithm of \cite{PZD:21:arXiv} are plotted against the number of qubits by utilizing $N_{\mathrm{tot}}=10^4$ to $10^9$ copies (starting from above) total. We can see that the median infidelities of our algorithm are better than the previously established technique for the same number of qubits and total measurements. } \label{fig3}
\end{figure}
We first compare our algorithm with the performance of improved estimation of five basis algorithm given in \cite{ZPD:19:PRA}. First, measurement in the computational basis gives the amplitude structure $\left|c_{i}\right|$ of the given unknown quantum state $\ket{\psi}=\sum_{i=0}^{d-1}c_{i}\ket{i}$. Then, we employ the permutation unitary on the given unknown quantum state such that
\begin{align}
	\ket{\psi_{p}}=U_{p}\ket{\psi}=\sum_{i=0}^{d-1}c_{\left[i\right]}\ket{i},
\end{align} 
where the amplitudes have descending order as $\left|c_{\left[ 0\right]}\right|\geq \left|c_{\left[ 1\right]}\right|\geq \cdots \geq \left|c_{\left[ d-1\right] }\right|$ and the permutation unitary
\begin{align}
	U_p = \sum_{i = 0}^{d-1}\ket{i}\bra{\left[ i\right]}.
\end{align}
We use \eqref{eq:off_diagonal} used in matrix-completion quantum state tomography to find other entries. To further improve the accuracy, we first find all entries of first column and row in the density matrix using our algorithm and construct the complete density matrix $\rho$ in a single step as follows
\begin{align}
	\hat{\rho}_{i,j}=\frac{\hat{\rho}_{1,i}\hat{\rho}_{j,1}}{\hat{\rho}_{1,1}}.
\end{align}	
 Second, we obtain the complex phase $e^{\dot{\iota}\hat{\phi}_{i}}$ of estimated state from our algorithm and construct the pure quantum state using these phases and amplitudes in \eqref{eq:diagonal} by
\begin{align}
	\ket{\hat{ \psi}_{p}}=\sum_{i=0}^{d-1}|c_{i}^{'}|e^{\dot{\iota}\hat{\phi}_{i}} \ket{i}.
\end{align}
By applying the inverse of unitary, we revert back to the desired quantum state
\begin{align}
	\ket{\hat{ \psi}}=U_{p}^{\dagger}\ket{\hat{ \psi}_{p}}.
\end{align}
We plot median infidelity of $10^{3}$ pure randomly generated states according to Haar measure by  employing  matrix-completion quantum state tomography algorithm with improved five basis method. Five basis uses two entangled basis to estimate the pure quantum state which has high gate errors and decoherence in practical settings. Using our algorithm, not only we overcome this deficiency for practical quantum computers but also surpass the accuracy in numerical simulations as demonstrated in  Figure~\ref{fig10}. 

To determine the performance of our procedure, we show the median infidelity of $10^{3}$ Haar-random pure quantum states on $N_{\mathrm{tot}}=10^4$ to $10^9$ copies as a function of the number of qubits in Figure~\ref{fig3}. For comparison, we also plot the median infidelity of scalable estimation that also uses the local bases as proposed in \cite{PZD:21:arXiv}. Figure~\ref{fig3} shows that our algorithm has significant improvement over the existing algorithm of local bases measurement-based state tomography. The performance increase becomes more prominent when the number of qubits is increased. 

	\begin{figure}[t!]
	\centering
	\includegraphics[width = 0.6 \textwidth]{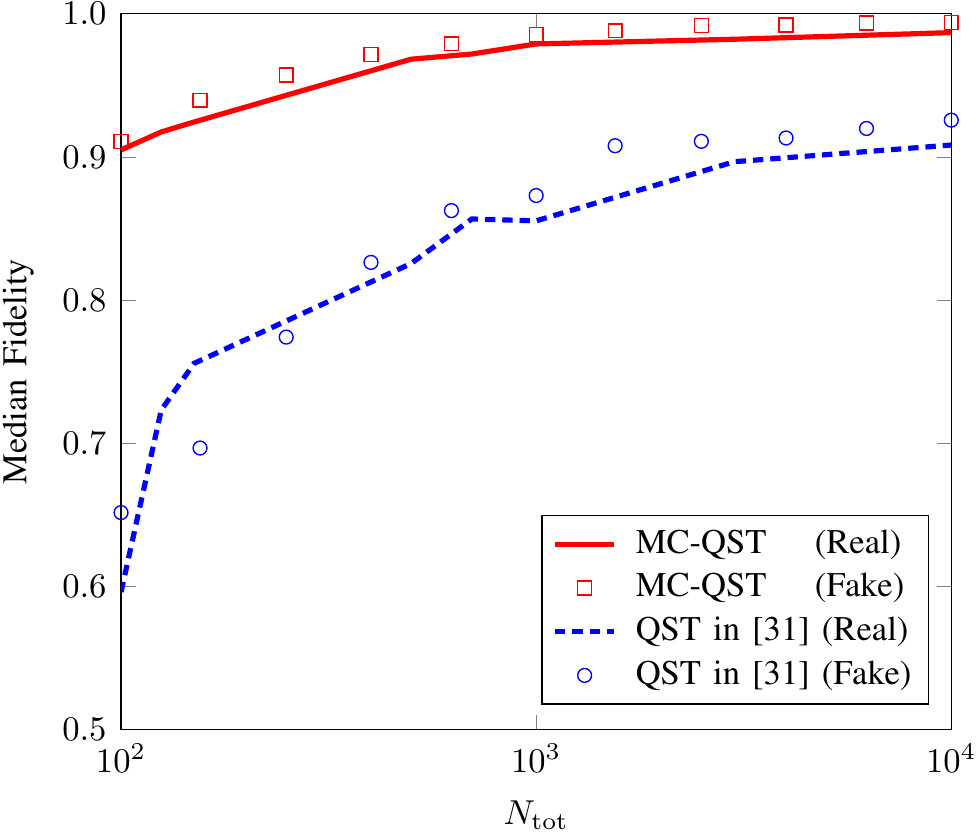}
	\caption{ Median infidelities of two-qubit Haar-random states as a function of total measurements. 
		The median values are calculated over $10^{2}$ states on real device and $10^3$ states on fake devices.  Graph shows a comparison of matrix-completion quantum state tomography (MC-QST) with the algorithm of \cite{PZD:21:arXiv}. We can clearly see that median fidelity of our algorithm outperforms the method \cite{PZD:21:arXiv}. } \label{fig4}
\end{figure}

We have also implemented our technique on a superconducting quantum computer to experimentally verify our results \cite{IBM:20:BE}. We use 7-qubit \texttt{ibmq-casablanca} device for these experiments. The device has average measurement and CNOT errors of $2.653e^{-2}$
and $8.374e^{-3}$ respectively. We also use the Fake-Casablanca which uses the same noise model as a real device to obtain more robust estimates by increasing the number of experiments. Figure~\ref{fig2} shows the circuit that we implement on the IBM quantum device where
\begin{align}
	H=\begin{bmatrix}
		1 &1 \\ 1&-1
	\end{bmatrix}, \hspace{0.25cm} \text{and} \hspace{0.25cm}
	S^{\dagger}=\begin{bmatrix}
		1 &0 \\ 0&e^{-\frac{\dot{\iota}\pi}{2}}
	\end{bmatrix}.
\end{align}
We produced $10^2$ Haar-random pure two-qubit states on the real device. For fake device, we utilized $10^3$ pure two-qubit states. We plot the median fidelity of our algorithm and scalable estimation of \cite{PZD:21:arXiv} in Figure~\ref{fig4}. Numerical results indicate that we can precisely benchmark the cloud-based IBM quantum devices.

Our algorithm is more effective when the density matrix is less sparse. To deal with the sparse matrix problem, we first measure in computational basis and see the structure of vanishing coefficients. Depending on the structure, we prepare the local unitary and apply it to the quantum state before measuring on a local basis.  Greenberger-Horne-Zeilinger (GHZ) state is one of the examples of this type of states. We prepare and measure three-qubits GHZ state in the computational basis on an IBM device.  We find out that it has only two dominant diagonal entries and the remaining entries are very small or zero. To overcome this issue, we prepare the local unitary of the form
\begin{align}
	V=H \otimes H \otimes H ,
\end{align}
and apply on the state
\begin{align}
	\ket{\psi'}=V\ket{\psi}.
\end{align} 
After estimating all the coefficients of the quantum state, we apply the inverse of the unitary matrix to get back the original prepared quantum state
\begin{align}
	\ket{\psi}=V^{\dagger}\ket{\psi'}.
\end{align}   

\begin{figure}[t!] 
	\centering     
	\subfigure[Re$\left(\rho\right)$]{\label{fig:a}\includegraphics[width=74mm]{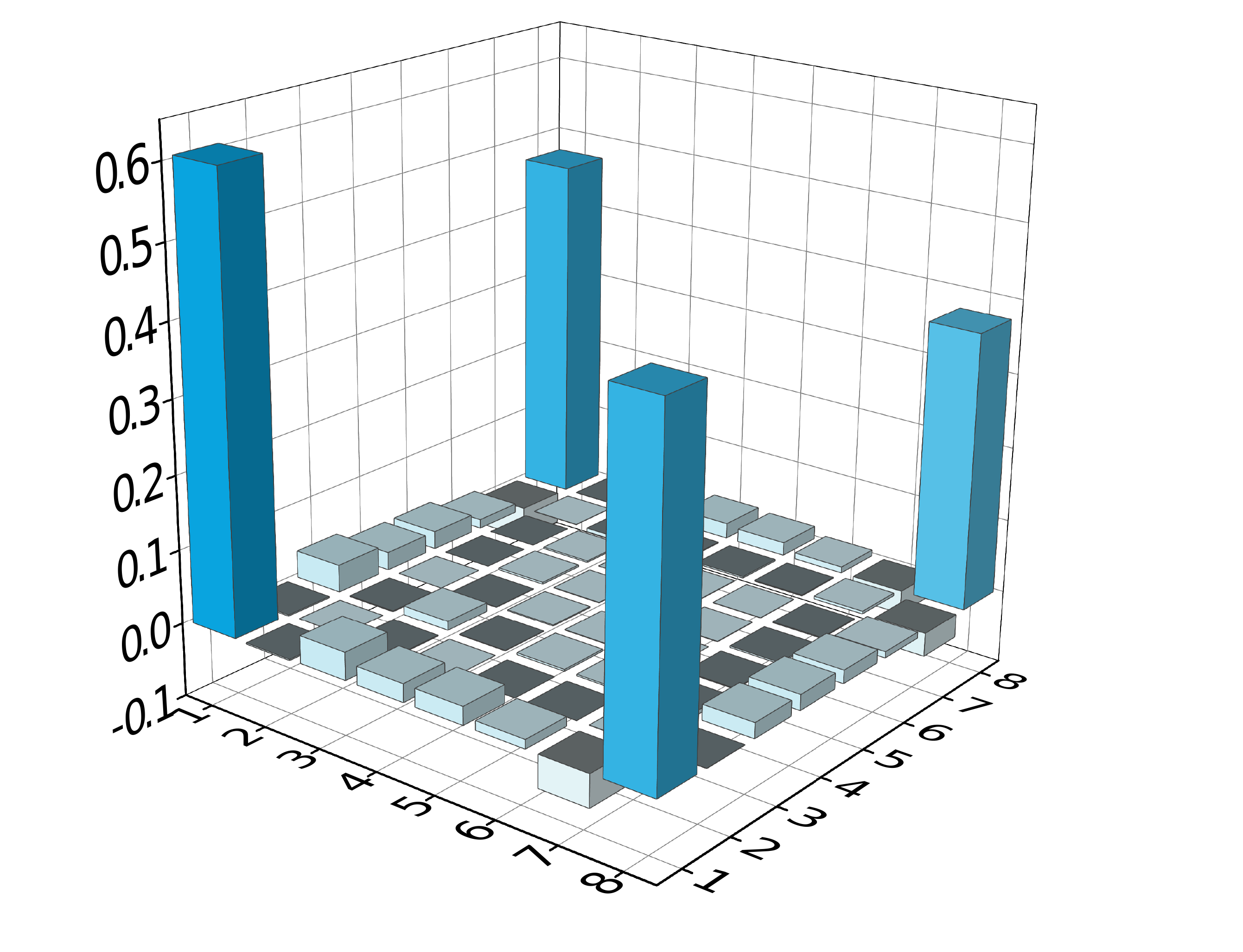}}
	\subfigure[Im$\left(\rho\right)$]{\label{fig:b}\includegraphics[width=74mm]{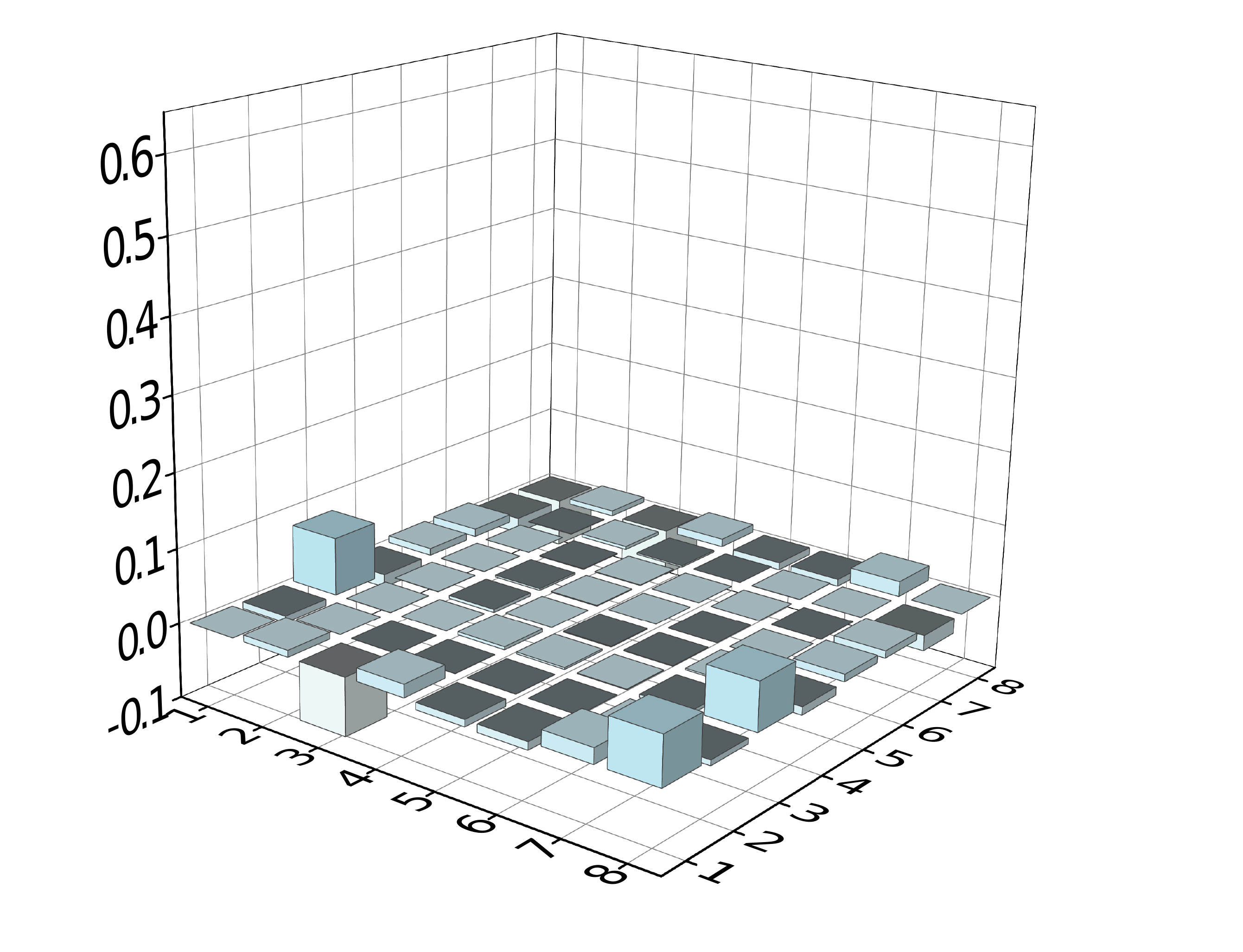}}
	\caption{  Reconstructing three-qubit GHZ state on a superconducting processor.  Real and imaginary parts of the matrix coefficients of the experimentally estimated GHZ state on \texttt{ibmq-casablanca} device are shown.   }\label{fig5}
\end{figure}

In Figure.~\ref{fig5}, we demonstrate the real and imaginary parts of the experimentally reconstructed GHZ state through our algorithm. We generate eleven GHZ states and select the median fidelity density matrix. For three qubits GHZ states, we can reconstruct the density matrix with $97\%$ median fidelity.

\section{Conclusion}\label{sec 4}
Entangling gates act as a bottleneck in the performance of accurate estimation and benchmarking tasks in NISQ devices.   
In this paper, we proposed an efficient algorithm for the estimation of pure states that is computationally efficient and requires only local measurements. Implementing the local Pauli bases on any quantum devices requires only local gates such as $H$ and $S$, which can be implemented with relatively more precision as compared to entangling gates. Furthermore, our algorithm reduces the circuit depth in real quantum devices, which is usually high when implementing entangling gates. This reduction of circuit depth can play a vital role where gate errors and decoherence rates are high. We observe the $97\%$ fidelity of qutrit GHZ state which is, to the best of our knowledge, the highest ever reported on superconducting quantum devices.   Experimental results demonstrate that our technique is extremely efficient for benchmarking of cloud-based NISQ computers available today. 

\section*{Acknowledgments}
We acknowledge the use of the IBM Q for this work. The views expressed are those
of the authors and do not reflect the official policy or position of IBM or the IMB Q team. This work was
supported by the National Research Foundation of Korea (NRF) grant funded by the Korea government
(MSIT) (No. 2019R1A2C2007037) and by the MSIT (Ministry of Science and ICT), Korea, under the ITRC
(Information Technology Research Center) support program (IITP-2021-0-02046) supervised by the IITP
(Institute for Information $\&$ Communications Technology Planning $\&$ Evaluation).
\bibliographystyle{unsrtnat}

\end{document}